\documentclass{article}

\usepackage{arxiv}

\usepackage[utf8]{inputenc}
\usepackage[T1]{fontenc}
\usepackage{hyperref}
\usepackage{url}

\usepackage{booktabs}
\usepackage{amsfonts}
\usepackage{nicefrac}
\usepackage{microtype}
\usepackage{amsmath}

\usepackage{lipsum}
\usepackage{wrapfig}
\usepackage{graphicx}
\usepackage{newfloat}
\usepackage{listings}
\usepackage{algorithmic}

\usepackage{subcaption}
\usepackage{tabularx}
\usepackage{multirow}
\usepackage{longtable}

\graphicspath{ {./images/} }

\title{Predicting Open Source Software Sustainability with Deep Temporal Neural Hierarchical Architectures and Explainable AI}

\author{
S M Rakib Ul Karim \\
Dept. of Electrical \& Computer Engineering\\
University of Missouri\\
Columbia, Missouri, United States \\
\texttt{skarim@missouri.edu} \\
\And
Wenyi Lu \\
Dept. of Computer Science \\
University of Missouri\\
Columbia, Missouri, United States \\
\texttt{wldh6@mail.missouri.edu} \\
\And
Enock Kasaadha \\
Dept. of Computer Science \\
University of Missouri\\
Columbia, Missouri, United States \\
\texttt{ekqkf@missouri.edu} \\
\And
Sean Goggins \\
Dept. of Electrical \& Computer Engineering\\
University of Missouri\\
Columbia, Missouri, United States \\
\texttt{gogginss@missouri.edu} \\
}

\begin{document}
\maketitle
\begin{abstract}
Open Source Software (OSS) projects follow diverse lifecycle trajectories shaped by evolving patterns of contribution, coordination, and community engagement. Understanding these trajectories is essential for stakeholders seeking to assess project organization and health at scale. However, prior work has largely relied on static or aggregated metrics, such as project age or cumulative activity, providing limited insight into how OSS sustainability unfolds over time. In this paper, we propose a hierarchical predictive framework that models OSS projects as belonging to distinct lifecycle stages grounded in established socio-technical categorizations of OSS development. Rather than treating sustainability solely as project longevity, these lifecycle stages operationalize sustainability as a multidimensional construct integrating contribution activity, community participation, and maintenance dynamics. The framework combines engineered tabular indicators with 24-month temporal activity sequences and employs a multi-stage classification pipeline to distinguish lifecycle stages associated with different coordination and participation regimes. To support transparency, we incorporate explainable AI techniques to examine the relative contribution of feature categories to model predictions. Evaluated on a large corpus of OSS repositories, the proposed approach achieves over 94\% overall accuracy in lifecycle stage classification. Attribution analyses consistently identify contribution activity and community-related features as dominant signals, highlighting the central role of collective participation dynamics.
\end{abstract}

\section{Introduction}
Open source software (OSS) has become the backbone of modern software development, powering critical infrastructure, enabling innovation, and fostering global collaboration. However, the sustainability of OSS projects remains a significant challenge, with estimates suggesting that over 80\% of OSS projects become inactive within a few years of inception \cite{robles2014floss,teixeira2017release}. The ability to anticipate and understand sustainability trajectories is therefore critical for developers, maintainers, funders, and organizations that depend on these ecosystems.

OSS sustainability encompasses multiple dimensions, including technical maintenance, community engagement, contributor retention, and long-term organizational viability. Much prior work has operationalized sustainability using static or aggregated indicators, such as code quality, documentation completeness, or developer counts \cite{capiluppi2007cathedral,mockus2002two}. While informative, these approaches offer limited insight into how sustainability unfolds over time, as they fail to capture the dynamic evolution of activity patterns, contributor behavior, and community interaction.

Recent advances in deep learning and temporal modeling have opened new opportunities for analyzing complex sequential data in software engineering contexts \cite{hoang2019deepjit}. Architectures such as recurrent neural networks, temporal convolutional networks, and Transformers have demonstrated strong capacity to model long-range dependencies and temporal structure in time-series data\cite{vaswani2017attention,bai2018empirical}. In OSS research, these methods have been applied to tasks such as defect prediction and project health assessment, suggesting their potential for more nuanced sustainability modeling \cite{hoang2019deepjit}.

However, the application of deep temporal models to OSS sustainability remains limited in scope and depth \cite{han2024sustainability}. Existing studies frequently frame sustainability as a binary outcome (e.g., active vs.\ inactive) or rely on coarse categorizations that overlook intermediate organizational regimes and lifecycle transitions \cite{kalliamvakou2014promises}. Moreover, many deep learning approaches emphasize predictive performance without providing interpretable explanations, constraining their usefulness for stakeholders who require actionable understanding rather than opaque predictions \cite{yin2021sustainability}.

Motivated by these gaps, this study addresses the following research questions: 
\begin{itemize}
    \item \textbf{RQ1:} To what extent can OSS sustainability stages be predicted from a fixed 24-month window of repository activity using temporal sequences and derived activity-based features?
    \item \textbf{RQ2:} Which categories of repository activity most strongly influence sustainability stage predictions, and how do these attribution patterns vary across modeling approaches and decision components?
\end{itemize}

To answer these questions, we explore and empirically validate a hierarchical neural modeling strategy that integrates Transformer-based temporal processing with feedforward neural modeling of engineered tabular features. Rather than relying on a single flat classifier, the approach decomposes the prediction task into specialized decision components designed to better capture heterogeneous lifecycle regimes and mitigate class imbalance. Explainable AI techniques are incorporated to systematically analyze how different categories of repository activity contribute to model decisions.

Our study makes four primary contributions. First, we show that OSS sustainability stages can be accurately predicted from recent activity histories when temporal and derived activity-based features are jointly modeled, substantially outperforming flat baselines. Second, our analyses reveal that recent contribution and community dynamics consistently dominate sustainability predictions, underscoring the importance of short- to mid-term activity patterns. Third, we introduce an attribution-based explainability framework that aggregates feature-level explanations into interpretable activity categories, enabling transparent and actionable insights into model behavior. Finally, we release an open-source implementation of the full pipeline to support reproducibility and future research. These contributions advance the study of OSS sustainability by demonstrating how temporal modeling and explainable learning can be combined to support scalable, transparent assessment of project health, with implications for maintainers, funding organizations, and platform-level ecosystem monitoring.


\section{Related Work}

\subsection{Open Source Software Sustainability and Prediction}
\label{rel:sub1}

Open source software (OSS) sustainability has become a central topic in software engineering research as OSS increasingly underpins critical digital infrastructure. Prior studies have sought to characterize, predict, and improve sustainability using empirical signals derived from repository activity, developer behavior, and project organization. This literature reveals steady methodological progress, alongside persistent conceptual and technical limitations.

Early work primarily relied on static or snapshot-based indicators of project health, such as contributor counts, commit volume, code churn, issue activity, and communication traces. For example, Mockus et al.~\cite{mockus2002two} analyzed developer participation and communication patterns in the Apache ecosystem, while Capiluppi and Michlmayr~\cite{capiluppi2007cathedral} examined project evolution through code churn and contributor turnover. Subsequent studies incorporated social network analysis to model collaboration structures and developer roles~\cite{joblin2017classifying}. More recent work has explored community engagement signals, including emoji usage and participation dynamics, as potential sustainability indicators~\cite{zhou2024emoji,weld2024making}. While foundational, these approaches largely capture sustainability through aggregated or cross-sectional views, limiting their ability to represent evolving project dynamics.

To address this limitation, later studies emphasized temporal and longitudinal modeling. Survival analysis and hazard-based models have been applied to study project abandonment and longevity~\cite{robles2014floss,teixeira2017release}, while machine learning and deep learning techniques have leveraged fine-grained repository traces for defect prediction and activity forecasting~\cite{hoang2019deepjit}. Lu et al.~\cite{lu2023team}, for example, examined OSS sustainability and team coordination using temporal metrics under external disruption. Although these methods incorporate time-dependent signals, they often focus on narrow metric subsets and operationalize sustainability as a single outcome (e.g., survival or inactivity), rather than a multidimensional organizational state.

Parallel work has aimed to support actionable sustainability assessment and decision making. Singhvi et al.~\cite{singhvi2024predicting} introduced the Eclipse Project Explorer (EPEX) to forecast project graduation using socio-technical networks, while Khan and Filkov proposed the ReACT framework and tool to operationalize evidence-based sustainability actions~\cite{singhvi2024predicting}. Despite their practical value, these systems rely heavily on predefined indicators and project-level summaries, offering limited insight into temporal transitions or intermediate organizational regimes.

Across this literature, several gaps remain. First, many studies rely on popularity-driven or partial proxies for sustainability (e.g., stars or contributor volume), which may obscure structural differences between projects. Second, temporal signals are often modeled in isolation from rich tabular summaries, limiting integration of short-term dynamics with longer-term trends. Third, severe class imbalance—particularly for mature or highly coordinated projects—is rarely addressed explicitly, reducing reliability for minority but socially important lifecycle regimes. Finally, although explainable AI (XAI) techniques are well established in machine learning~\cite{lundberg2017unified,ribeiro2016should}, their application to OSS sustainability remains limited, with few studies offering transparent, model-level explanations to support actionable interpretation~\cite{yeow2025predicting,myakala2025explainable}.

These limitations motivate the need for more nuanced representations of OSS sustainability that move beyond single-metric or longevity-based definitions. Lifecycle-stage frameworks grounded in socio-technical organization~\cite{eghbal2016roads} provide a richer lens for capturing how repositories evolve across distinct coordination and participation regimes. By integrating temporal modeling, multidimensional activity features, hierarchical classification, and modern XAI techniques, this work aims to advance scalable and interpretable OSS sustainability prediction.

\subsection{Advanced Computational Methods for Analyzing Temporal OSS Activity}

Recent advances in computational methods have significantly expanded the analytical toolkit available for studying OSS activity and evolution. In response to the limitations identified in prior sustainability research, such as reliance on static metrics, limited temporal modeling, and lack of interpretability, researchers have increasingly turned to machine learning, deep learning, and structured learning paradigms to better capture the complexity of OSS ecosystems. This section reviews key methodological developments that inform our approach.

\paragraph{Mining Software Repositories with Machine Learning}
Mining Software Repositories (MSR) research has long leveraged machine learning techniques to analyze large-scale repository data for tasks such as defect prediction, maintainability assessment, developer recommendation, and software reuse forecasting \cite{li2025comparison,shi2024time}. Prior work demonstrates that activity traces extracted from version control systems, issue trackers, and pull request logs provide valuable predictive signals for software quality and maintenance outcomes. However, many MSR-based models rely on aggregated or short-horizon features and emphasize predictive performance over interpretability, limiting their ability to explain longitudinal project dynamics or organizational transitions \cite{wang2024tglite,infant2025explainable}.

\paragraph{Deep Learning for Software Engineering}
Building on MSR foundations, deep learning approaches have been widely adopted for software engineering tasks, including defect prediction, code analysis, and maintenance forecasting \cite{velarde2022open,haldar2024interpretable,olawumi2025ai}. Recurrent neural networks, convolutional architectures, and attention-based models have demonstrated improved performance by learning hierarchical and nonlinear representations of software activity \cite{amoui2009temporal}. Despite these advances, prior deep learning studies in software engineering often treat projects as flat prediction instances and focus on single outcomes, such as defects or failures, without explicitly modeling lifecycle structure or addressing severe class imbalance across project categories \cite{bolon2024review}.

\paragraph{Temporal Neural Networks and Time Series Prediction}
Temporal modeling, as a subset of Deep Learning, has emerged as a critical capability for analyzing sequential and behavioral data \cite{smith2020temporal}. Long Short-Term Memory (LSTM) networks, temporal convolutional networks, and Transformer architectures have proven effective in capturing long-range dependencies and complex temporal patterns across diverse application domains. In software engineering, temporal neural networks have been applied to tasks such as defect prediction, activity forecasting, and maintenance planning, demonstrating that sequence-aware models outperform static approaches when longitudinal data are available \cite{wang2024tglite,jin2025impact}. Transformer-based models, in particular, offer flexible attention mechanisms that enable selective focus on salient time periods, motivating their adoption for modeling extended OSS activity histories \cite{li2025comparison}.

\paragraph{Hierarchical Classification and Class Imbalance Handling}
In OSS sustainability modeling, severe class imbalance across lifecycle stages poses a substantial challenge for multi-class prediction, often degrading performance on minority but structurally important categories \cite{bolon2024review}. Hierarchical learning strategies have been proposed to address challenges associated with complex label structures and severe class imbalance. By decomposing multi-class prediction tasks into structured decision hierarchies, hierarchical classifiers enable specialized models to focus on separable subsets of classes, improving minority-class recognition and reducing bias toward dominant categories \cite{germanos2024change,yunita2025performance}. Prior work across domains, including imbalanced regression, text classification, and multi-stage prediction pipelines, demonstrates that hierarchical decomposition can improve robustness and interpretability when class boundaries are asymmetric or overlapping, as is often the case in OSS lifecycle categorization \cite{jha2019deep,reza2023analyzing,bolon2024review}.

\paragraph{Ensemble Methods and Multi-Stage Learning}
Ensemble learning and multi-stage architectures further extend hierarchical approaches by combining multiple specialized models through confidence-aware routing or adaptive selection mechanisms \cite{tarwani2016predicting,haldar2024interpretable,yunita2025performance}. Rather than relying on a single global classifier, ensemble systems leverage complementary strengths of diverse learners, dynamically selecting predictions based on confidence estimates or error patterns. Such strategies have been shown to improve performance and stability in settings characterized by heterogeneous data distributions, noisy signals, and imbalanced classes, conditions that closely resemble real-world OSS ecosystems \cite{kaufman2024analyzing,nayak2024transformer,lawrynczuk2025lstm}.

\paragraph{Explainable AI for Software Engineering}
As predictive models grow in complexity, XAI techniques have become increasingly important for enabling transparency and stakeholder trust in software engineering applications. Feature attribution methods such as SHAP and Integrated Gradients have been applied to defect prediction, feature selection, and decision-support systems, offering insight into model behavior and key predictive signals. However, most existing XAI applications in software engineering focus on static or tabular inputs, with limited attention to explaining temporal representations or sequential decision processes. In the context of OSS sustainability, there remains a notable gap in integrating temporal deep learning models with interpretable, lifecycle-aware explanations \cite{infant2025explainable}.

\subsection{Synthesis and Motivation}
Taken together, these research streams highlight both the promise and the limitations of existing computational approaches for OSS sustainability analysis. Prior work demonstrates the value of temporal modeling, deep learning, hierarchical classification, and ensemble methods, yet these techniques are rarely integrated within a unified framework that simultaneously addresses sustainability prediction, class imbalance, temporal dynamics, and interpretability. Motivated by the gaps identified before, our work synthesizes these advances by combining Transformer-based temporal modeling, hierarchical and confidence-based classification, and modern XAI techniques to provide a scalable and interpretable approach for predicting and explaining OSS lifecycle stages.

\section{Methodology}
\label{sec:methodology}
We develop a hierarchical temporal pipeline to predict OSS sustainability stages from 24-month repository activity 
First, we construct standardized time-series and engineered tabular features, then route projects through Stage-1 (binary gate) and Stage-2 (Heavy Transformer + MLP or Light MLP), with a Club-Fed expert for minority classes. Training uses stratified group-aware splits, focal loss, class-weighted sampling, and calibration; evaluation reports accuracy, macro/weighted F1, and balanced accuracy. We further interpret decisions via SHAP/Integrated Gradients and targeted ablations \cite{lundberg2017unified, sundararajan2017axiomatic}.

\subsection{Dataset Construction and Preprocessing}

\subsubsection{Data Collection and Label Definition}

We construct a dataset of OSS repositories using monthly-aggregated activity logs from a random sample of 20k repositories, spanning a fixed 24-month observation window (June 2022 to June 2024). For each repository, the dataset includes 20 base metrics (see details in Appendix A) capturing multiple dimensions of development and community activity. The metrics are organized into five conceptual categories: contribution activity, community dynamics, issue responsiveness, pull request quality assurance, and release evolution.


All metrics are aggregated at the monthly level to preserve temporal dynamics while ensuring robustness to short-term noise. Each repository is assigned a target label representing its lifecycle stage, based on the socio-technical framework proposed by Eghbal~\cite{eghbal2016roads}. This framework characterizes OSS projects according to patterns of contributor participation, governance structure, and community visibility, rather than project age alone. Due to extremely limited data availability, we exclude the \textit{stadium} stage, which represents highly commercialized projects with professionalized maintenance and exceptionally large user bases. The remaining four stages are defined as follows:

\begin{itemize}
    \item \textbf{federation} (\textit{Collaborative Network}): Projects supported by multiple organizations or coordinated groups, typically exhibiting federated governance and shared maintenance responsibilities. These projects are characterized by a large contributor base (unique contributors $>$ 75) and high external visibility (stargazers $>$ 1{,}000), with a ratio of stargazers to unique contributors greater than 2. The average repository longevity under this label is 6.3 years.
    \item \textbf{club} (\textit{Exclusive Community}): Projects maintained by tightly knit expert communities with relatively high barriers to entry. While contributor counts are comparable to federated projects (unique contributors $>$ 75), these repositories exhibit lower external visibility, reflected by a stargazers-to-contributors ratio below 2. The average repository longevity under this label is 5.2 years.
    \item \textbf{contribMid} (\textit{Moderate Contributor Engagement}): Projects with moderate and balanced contributor participation, representing a transitional regime between personal and highly collaborative development. These projects typically involve between 6 and 75 unique contributors. The average repository longevity under this label is 2.8 years.
    \item \textbf{toy} (\textit{Minimal Activity}): Small-scale, often personal projects with limited activity and minimal community engagement. These repositories are characterized by fewer than 6 unique contributors and fewer than 100 stargazers. The average repository longevity under this label is 1.5 years.
\end{itemize}

\subsubsection{Temporal Feature Engineering}

For each repository $i$, we construct a temporal feature sequence $\mathbf{X}_i \in \mathbb{R}^{T \times F}$, where $T = 24$ denotes the number of monthly time steps and $F = 20$ is the number of base activity features. To ensure comparability across repositories and features, all temporal features are standardized using z-score normalization (see Section ~\ref{subsec:z-score-norm}). To explicitly model short-term temporal dynamics, we additionally compute first-order difference features that represent month-over-month changes in each base metric. These delta features capture local growth, decline, and volatility patterns that may not be evident from absolute activity levels alone. The final temporal representation concatenates the normalized activity levels with their corresponding temporal differences, yielding a sequence $\mathbf{X}_i = [\mathbf{X}_i^{\text{levels}}, \Delta \mathbf{X}_i] \in \mathbb{R}^{T \times 2F}.$ This combined representation preserves both long-term activity intensity and short-term temporal variation, enabling downstream models to jointly reason about sustained engagement and recent behavioral shifts \cite{box2015time}.

\subsubsection{Tabular Feature Engineering}
To complement the full temporal sequences, we engineer a set of tabular features designed to summarize recent project activity and support efficient classification in models that do not directly process sequential inputs. These features are derived from the final $K$ months of each project’s temporal history, where $K = 6$ is selected based on validation performance to balance sensitivity to recent behavior with robustness to short-term noise.

For each base feature $F$, we compute three groups of derived statistics over the last $K$ months: (1) \textbf{Recent values}: The raw feature values from the most recent window, $\mathbf{x}_{i,f}^{(t-K+1)}, \ldots, \mathbf{x}_{i,f}^{(t)}$; (2) \textbf{Statistical summaries}: The mean ($\mu_f$), standard deviation ($\sigma_f$), minimum ($\min_f$), maximum ($\max_f$), and linear trend coefficient (slope) computed over the same window; (3) \textbf{Delta statistics}: The mean ($\mu_{\Delta f}$) and standard deviation ($\sigma_{\Delta f}$) of the corresponding first-order temporal differences, capturing short-term variability and directional change.

In addition, we include a small set of cross-feature aggregations that reflect higher-level repository characteristics, including total activity ($\texttt{commit\_count} + \texttt{issues\_count}$), contributor engagement ratio ($\texttt{repeat\_contributors} / \texttt{committer\_count}$), and review efficiency ($\texttt{pr\_acceptance\_rate} \times \texttt{pr\_review\_duration}$).

The resulting tabular representation is $\mathbf{x}_i \in \mathbb{R}^{D}$, where
$D = F \times (K + 5 + 2) + C$, with $F = 20$ base features and $C = 3$ cross-feature aggregations, yielding approximately $D \approx 200$ dimensions. All tabular features are standardized using z-score normalization and are used as input to the MLP components in Stage-1, Stage-2 Light, and the Club-Fed Expert. This representation provides a compact yet interpretable summary of recent project health, consistent with established feature-engineering practices for high-dimensional predictive modeling \cite{guyon2003introduction}. More details regarding the definitions of preprocessing methods used in this section can refer to Appendix B. 

\subsection{Hierarchical Model Architecture}

\begin{figure*}[t]
  \centering
  \includegraphics[width=0.75\textwidth]{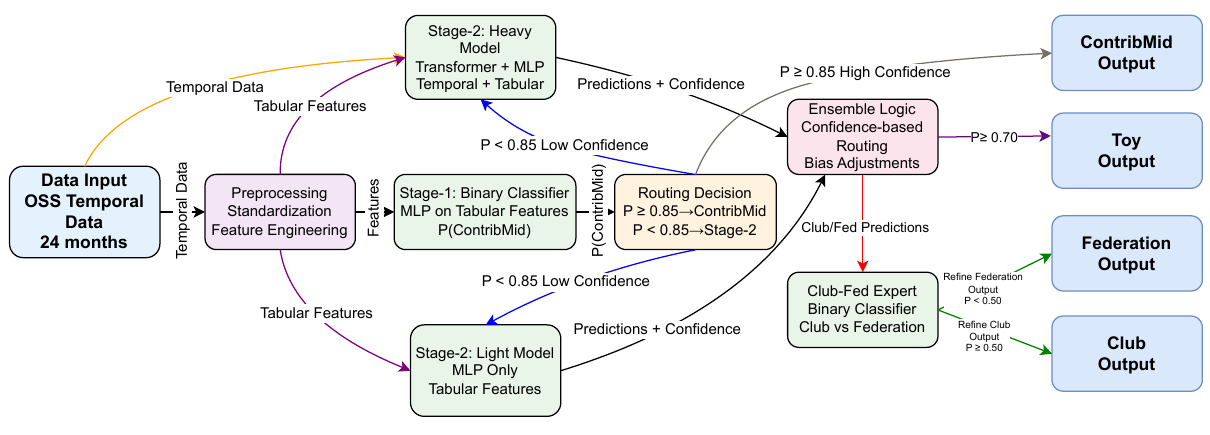}
  \caption{Overview of the hierarchical OSS lifecycle prediction pipeline. Temporal and tabular features are processed through staged classifiers with confidence-based routing to produce final lifecycle stage predictions.}
  \label{fig:architecture_hierarchy}
\end{figure*}

To address the multi-class OSS sustainability stage classification problem under severe class imbalance and overlapping behavioral regimes, we propose a hierarchical, multi-stage classification architecture inspired by prior work on hierarchical learning for imbalanced data \cite{silla2011survey}. The core intuition is to decompose a complex four-class prediction task into a sequence of simpler, task-specific decisions, allowing specialized models to focus on separable subsets of sustainability stages. Figure~\ref{fig:architecture_hierarchy} provides a high-level overview of the proposed pipeline. The architecture integrates both temporal and tabular representations of repository activity and routes projects through multiple decision stages based on prediction confidence. Figure~\ref{fig:model_architectures} further details the internal architectures of the four independently trained models that constitute the pipeline.

\subsubsection{Stage-1: Binary Gating for ContribMid Detection}

The first stage acts as a binary gating mechanism that distinguishes \texttt{contribMid} projects from all other lifecycle stages. As shown in Figure~\ref{fig:architecture_hierarchy}, a multilayer perceptron (MLP) operating on tabular features estimates the probability $P(\text{contribMid})$ for each project. Projects with $P(\text{contribMid}) \geq 0.85$ are directly classified as \texttt{contribMid} and exit the pipeline, reflecting the relative separability and prevalence of this class. Projects below this confidence threshold are routed to Stage-2 for finer-grained analysis. The Stage-1 classifier is implemented as a tabular MLP with input dimensionality $D \approx 200$, three hidden layers (256, 128, and 64 units), ReLU activations, and a sigmoid output neuron \cite{goodfellow2016deep}. The threshold of 0.85 is selected based on validation-set performance to balance precision and recall while minimizing downstream error propagation.

\subsubsection{Stage-2: Parallel Multi-class Classification}

Projects not confidently classified as \texttt{contribMid} proceed to Stage-2, which consists of two parallel multi-class classifiers: a Heavy model and a Light model. This design enables a trade-off between expressive temporal modeling and computational efficiency, while also providing complementary perspectives on project behavior.

\paragraph{Heavy Model (Transformer + MLP)}
The Heavy model jointly processes temporal and tabular features. A Transformer encoder ingests the temporal sequence $\mathbf{X}_i \in \mathbb{R}^{24 \times 40}$, where 40 features correspond to concatenated activity levels and first-order differences. The Transformer comprises six layers with four attention heads, an embedding dimension of 192, and sinusoidal positional encoding \cite{vaswani2017attention}. The resulting temporal representation is concatenated with the tabular feature embedding produced by a two-layer MLP (128 and 64 units), and the combined representation is passed to a softmax output layer for three-way classification among \texttt{club}, \texttt{federation}, and \texttt{toy}. This model is designed to capture long-range temporal dependencies and nuanced behavioral transitions that may be critical for distinguishing structurally similar lifecycle stages.


\paragraph{Light Model (MLP-only Baseline)}
In parallel, the Light model operates exclusively on tabular features and consists of a three-layer MLP with 256, 128, and 64 hidden units and ReLU activations \cite{goodfellow2016deep}. Although less expressive than the Heavy model, this architecture provides a strong baseline that emphasizes recent activity summaries and enables efficient inference when temporal signals are weak or noisy.

\paragraph{Confidence-based Ensemble Routing}

Both Stage-2 models output class probability vectors along with associated confidence scores. 

Following Figure~\ref{fig:architecture_hierarchy}, we select the prediction from the model with higher confidence:
\begin{equation}
c_i^{\text{heavy}} = \max(\hat{\mathbf{y}}_i^{\text{heavy}}), \quad
c_i^{\text{light}} = \max(\hat{\mathbf{y}}_i^{\text{light}}),
\end{equation}
\begin{equation}
\hat{\mathbf{y}}_i^{\text{stage2}} =
\begin{cases}
\hat{\mathbf{y}}_i^{\text{heavy}} & \text{if } c_i^{\text{heavy}} \geq c_i^{\text{light}}, \\
\hat{\mathbf{y}}_i^{\text{light}} & \text{otherwise}.
\end{cases}
\end{equation}

Predictions classified as \texttt{toy} with confidence $ \geq 0.70$ are accepted as final outputs. Predictions assigned to \texttt{club} or \texttt{federation}, or those with lower confidence, are routed to a specialized expert classifier for further refinement.

\subsubsection{Club-Fed Expert for Minority Class Refinement}

To improve discrimination between the two most confusable and underrepresented classes, \texttt{club} and \texttt{federation}, we introduce a dedicated binary expert model. This Club-Fed Expert is an independent tabular MLP with two hidden layers (128 and 64 units) and a sigmoid output, trained exclusively to distinguish between these two classes.

The expert overrides the Stage-2 prediction according to:
\begin{equation}
\hat{y}_i^{\text{final}} =
\begin{cases}
\text{club} & \hat{y}_i^{\text{cf}} \geq \tau_{\text{cf}}, \\
\text{federation} & \hat{y}_i^{\text{cf}} < \tau_{\text{cf}},
\end{cases}
\end{equation}
where $\tau_{\text{cf}} = 0.5$ is optimized on the validation set \cite{zhou2012ensemble, ma2018model}.

More detailed architecture for each of the four models can be found in the \ref{fig:model_architectures} within Appendix C, along with mathematical formulas that integrate the models into the hierarchy architecture.


\subsection{Training Procedure}

All models in the hierarchical pipeline are trained independently on task-specific datasets derived from the same OSS repository corpus. To address substantial class imbalance, particularly for the \texttt{club} and \texttt{federation} stages, we employ a combination of inverse-frequency class weighting, weighted mini-batch sampling, and Synthetic Minority Oversampling Technique (SMOTE) for minority classes \cite{chawla2002smote}. For multi-class tasks, focal loss with class weights is used to emphasize hard-to-classify instances and reduce bias toward dominant classes.

Model optimization is performed using the Weighted Adam optimizer with weight decay to promote generalization \cite{loshchilov2017decoupled}. Learning rates are selected separately for each model based on validation performance, reflecting differences in task complexity. Training uses a fixed batch size of 256 for computational efficiency, with early stopping applied on validation accuracy (patience of six epochs). To improve stability and reduce overfitting, we apply batch normalization, dropout ($p = 0.1$), and gradient clipping with a maximum norm of 1.0 \cite{srivastava2014dropout}.

Hyperparameters are selected via grid search on a held-out validation split (20\% of the training data), using balanced accuracy and F1-score as selection criteria due to class imbalance. Search ranges include learning rates between $10^{-5}$ and $10^{-3}$, dropout rates in $[0.1, 0.3]$, hidden layer widths between 64 and 256, and Transformer depths and attention heads in the range of 4 to 8 for the Heavy model. Final configurations are chosen through iterative validation guided by prior literature and empirical performance. A summary of the selected hyperparameters for all models is reported in Table~\ref{tab:all-hyper} within Appendix D.

\subsection{Model Evaluation}

We evaluate the proposed hierarchical architecture using a temporally grounded and leakage-aware experimental setup. Projects are split at the repository level into training (70\%), validation (15\%), and test (15\%) sets using time-aware stratification to preserve chronological order and prevent future information from influencing past predictions. All reported results are computed on the held-out test set using the complete hierarchical pipeline.

Model performance is assessed using standard multi-class classification metrics, including accuracy, balanced accuracy, and macro and weighted F1-scores, to account for class imbalance across lifecycle stages. In addition to overall performance, we report per-class precision, recall, and F1-scores, and analyze confusion matrices to identify systematic misclassifications, particularly among structurally similar and minority classes such as \texttt{club} and \texttt{federation}. Decision thresholds for binary classification and routing are optimized on the validation set to maximize the F1-score. More detailed definition of the metrics used for evaluation can be found in Appendix E. 



\subsection{Explainability and Attribution Analysis}

To interpret model behavior and provide actionable insights into OSS sustainability stage predictions, we conduct a comprehensive explainability analysis spanning feature attribution, component-level assessment, and model-specific interpretability techniques. Given the hierarchical and hybrid nature of the proposed architecture, explainability is applied at both the individual model level and the full pipeline level.

\paragraph{Attribution Methods by Model Architecture}
For models operating on temporal sequences, we apply \textbf{Integrated Gradients (IG)} to the Stage-2 Heavy model. IG is a gradient-based attribution method designed for deep neural networks with structured inputs, and is particularly effective for sequential and high-dimensional representations. IG computes feature attributions by integrating gradients along a straight-line path from a baseline input to the observed input, yielding stable and path-aware importance estimates for temporal features.

For models operating on engineered tabular features, including Stage-1, Stage-2 Light, and the Club-Federation Expert, we apply \textbf{SHAP (SHapley Additive exPlanations)}. SHAP is grounded in cooperative game theory and provides additive feature attributions that account for feature interactions and correlated inputs. This makes SHAP well-suited for multilayer perceptrons (MLPs) trained on aggregated statistics, where individual feature contributions are not temporally ordered but may interact nonlinearly.

\paragraph{Feature-Level Attribution Computation}
For each trained model, we compute per-instance attribution scores for all input features using the corresponding XAI method (IG or SHAP). These scores quantify the contribution of each feature to the model’s predicted class probability. To facilitate comparability across samples and models, attribution magnitudes are converted to absolute values and normalized within each model.

\paragraph{Semantic Feature Categorization}
To enable higher-level interpretation, individual features are grouped into five semantically meaningful OSS activity categories:
\emph{Contribution Activity}, \emph{Issue \& Maintenance Responsiveness}, \emph{Community Dynamics}, \emph{Pull Request Quality Assurance}, and \emph{Release \& Evolution Metrics}.  
Each engineered feature is assigned to exactly one category based on its operational definition.

\paragraph{Category-Level Attribution Aggregation}
Category importance scores are computed by aggregating feature-level attributions within each category. Specifically, for category $c$, importance is defined as:
\[
I_c = \frac{1}{|F_c|} \sum_{f \in F_c} \mathbb{E}\left[ \left| a_f \right| \right],
\]
where $F_c$ denotes the set of features belonging to category $c$ and $a_f$ represents the attribution score of feature $f$. Expectations are taken over all test instances. This aggregation yields normalized, category-level importance scores that are directly comparable across models and decision stages.

\paragraph{Temporal Attribution Analysis}
For the Stage-2 Heavy Transformer, IG attributions are further analyzed across time steps to examine temporal importance patterns. Feature attributions are aggregated over feature dimensions within each month, enabling assessment of how predictive influence varies across the 24-month observation window. In addition, Transformer attention weights are inspected to corroborate attribution-based temporal patterns.

\paragraph{Ablation-Based Validation}
To validate that attribution scores reflect meaningful and causal contributions rather than spurious correlations, we conduct targeted feature ablation experiments. Entire feature categories are removed from the input space, models are retrained, and the resulting performance degradation is measured. Consistency between attribution rankings and ablation-induced performance drops is used as evidence of attribution robustness. More detailed definition of the metrics used for evaluation can be found in Appendix F. 



\section{Results}

\subsection{Predictive Performance of OSS Sustainability Stage Classification}

To answer our first research question, this section reports the empirical performance of the proposed hierarchical temporal pipeline for OSS sustainability stage prediction. We first compare the pipeline against multiple baseline architectures, then analyze class-wise behavior, and error patterns.

\paragraph{Overall Predictive Performance}
\label{sec:overall_comparison}

Table~\ref{tab:overall_comparison} summarizes overall classification performance across models. The proposed hierarchical pipeline substantially outperforms all baselines on every reported metric, achieving 94.08\% Accuracy, 79.42\% Balanced Accuracy, 78.96\% Macro F1, and 94.16\% Weighted F1. The gains in Macro F1 and Balanced Accuracy are particularly notable, as these metrics emphasize performance on minority classes and are less influenced by class prevalence. In contrast, baseline models, both tabular and sequence-based, exhibit limited ability to generalize beyond the dominant sustainability stages. These results demonstrate that sustainability (lifecycle) stages can be predicted with high accuracy from a fixed 24-month observation window when temporal activity histories and derived activity-based features are jointly modeled within a structured prediction framework.

\begin{table}[h]
\centering
\caption{Overall performance comparison across models. Baselines include a Multilayer Perceptron (MLP), Gated Recurrent Unit network (GRU), Temporal Convolutional Network (TCN), one-dimensional Convolutional Neural Network (1D CNN), CNN-LSTM hybrid model, and an Autoencoder-based classifier. The proposed Hierarchical Pipeline outperforms all baselines across Accuracy, Balanced Accuracy, Macro F1, and Weighted F1.}
\label{tab:overall_comparison}
\scriptsize
\setlength{\tabcolsep}{3.5pt}
\renewcommand{\arraystretch}{1.05}
\begin{tabularx}{\columnwidth}{l *{4}{>{\centering\arraybackslash}X}}
\toprule
\textbf{Model} & \textbf{Accuracy (\%)} & \textbf{Balanced Acc. (\%)} & \textbf{Macro F1 (\%)} & \textbf{Weighted F1 (\%)} \\
\midrule
MLP                 & 52.89 & 56.15 & 46.72 & 52.98 \\
GRU                 & 50.73 & 56.71 & 44.94 & 49.47 \\
1D CNN              & 42.54 & 49.87 & 41.71 & 48.83 \\
Autoencoder         & 37.92 & 43.74 & 36.63 & 44.64 \\
TCN                 & 32.80 & 50.14 & 29.46 & 19.36 \\
CNN--LSTM           & 17.69 & 33.01 & 18.33 & 19.96 \\
\midrule
\textbf{Hierarchical Pipeline (ours)} 
                    & \textbf{94.08} 
                    & \textbf{79.42} 
                    & \textbf{78.96} 
                    & \textbf{94.16} \\
\bottomrule
\end{tabularx}
\end{table}

\paragraph{Per-Class Performance and Error Patterns}

\begin{figure}[t]
  \centering
  \includegraphics[width=0.\columnwidth]{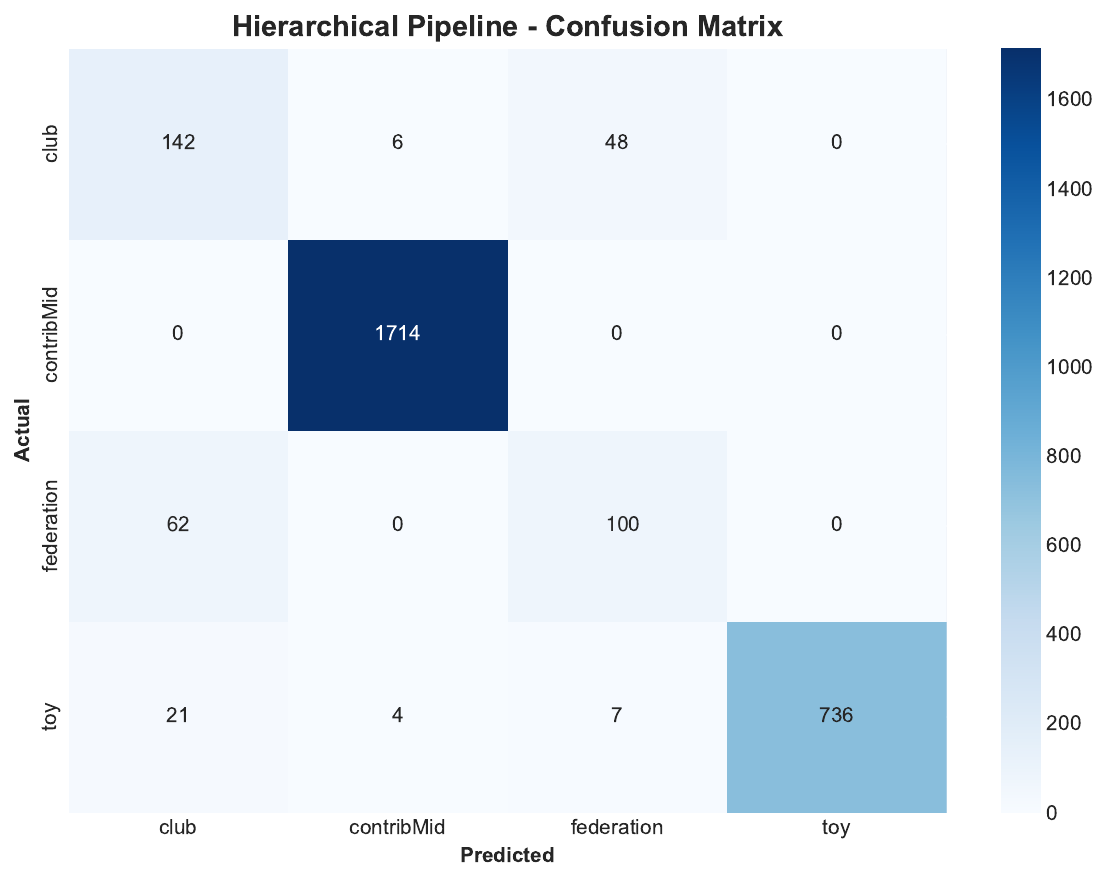}
  \caption{Confusion matrix for Hierarchical Pipeline predictions across four sustainability stages.}
  \label{fig:confusion_matrix}
\end{figure}

Table~\ref{tab:per_class} reports class-level precision, recall, and F1-scores for the hierarchical pipeline. The model achieves near-perfect performance for the \textbf{contribMid} and \textbf{toy} stages, with F1-scores exceeding 97\%, reflecting strong separability for projects with moderate engagement and minimal activity regimes. Performance is lower for the minority stages \textbf{club} and \textbf{federation}, with F1-scores of 64.40\% and 53.87\%, respectively. The resulting disparity between macro-averaged and weighted metrics reflects the underlying class imbalance in the dataset rather than systematic model failure. As shown in Figure~\ref{fig:confusion_matrix}, most misclassifications occur between the structurally similar \textbf{club} and \textbf{federation} stages, indicating that errors are concentrated among closely related lifecycle regimes rather than between dissimilar classes.

\begin{table}[h]
\centering
\caption{Per-class precision, recall, and F1-score for the Hierarchical Pipeline.}
\label{tab:per_class}
\footnotesize
\setlength{\tabcolsep}{3.5pt}
\renewcommand{\arraystretch}{1.05}
\begin{tabularx}{\columnwidth}{l *{4}{>{\centering\arraybackslash}X}}
\toprule
\textbf{Sustainability Stage} & \textbf{Precision} & \textbf{Recall} & \textbf{F1-Score} & \textbf{Support} \\
\midrule
club         & 0.5796 & 0.7245 & 0.6440 & 196 \\
contribMid   & 0.9942 & 1.0000 & 0.9971 & 1714 \\
federation   & 0.5926 & 0.4938 & 0.5387 & 162 \\
toy          & 1.0000 & 0.9583 & 0.9787 & 768 \\
\midrule
Macro Avg    & 0.7916 & 0.7942 & 0.7896 & 2840 \\
Weighted Avg & 0.9442 & 0.9408 & 0.9416 & 2840 \\
\bottomrule
\end{tabularx}
\end{table}

\paragraph{Effects of Class Imbalance}

Performance differences across stages are largely driven by pronounced class imbalance. The \textbf{contribMid} class accounts for 60.25\% of all samples, followed by \textbf{toy} at 27.04\%, while \textbf{club} and \textbf{federation} together comprise less than 13\% of the dataset. Limited representation reduces the amount of training signal available for minority stages and constrains recall, particularly for \textbf{federation}, which exhibits the lowest recall (0.4938).

The hierarchical structure of the pipeline further amplifies this effect: misclassifications at earlier decision stages can propagate downstream, disproportionately affecting minority classes. Nevertheless, the high weighted F1-score (0.9416) indicates that the model delivers stable and reliable predictions at scale, which is particularly important for ecosystem-level sustainability monitoring where majority-class accuracy is critical (e.g., large-population sustainability monitoring). Overall, the results provide strong empirical support for \textbf{RQ1}, indicating that OSS sustainability stages can be accurately inferred from recent repository activity when temporal and activity-based features are effectively integrated.

\subsection{Attribution-Based Analysis of Activity Category Contributions}

\begin{figure}
  \centering
  \includegraphics[width=0.7\columnwidth]{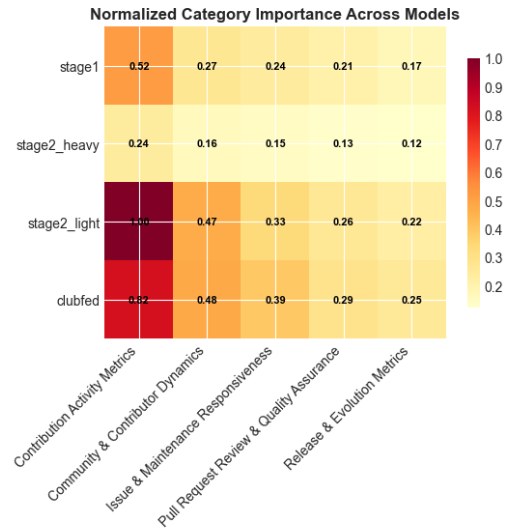}
  \caption{Normalized category importance heatmap showing relative influence of feature categories (rows) across different model architectures (columns), with values normalized to [0,1] scale.}
  \label{fig:heated_map}
\end{figure}

To answer \textbf{RQ2}, which categories of repository activity most strongly influence sustainability stage prediction, and how these attribution patterns vary across modeling approaches and decision components, we conduct a comprehensive explainability analysis using SHAP, Integrated Gradients, Transformer attention inspection, and targeted feature ablation. Figure~\ref{fig:heated_map} summarizes normalized category-level importance across all models in the hierarchical pipeline.

\begin{figure}
  \centering
  \includegraphics[width=0.7\columnwidth]{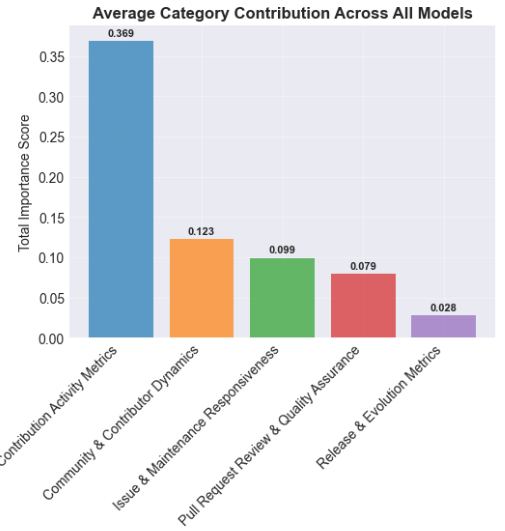}
  \caption{Average category contribution bar chart displaying mean total Top Temporal Features by Importance Scores aggregated across all models, sorted by descending influence.}
  \label{fig:bar-plot}
\end{figure}

\paragraph{Cross-Model Category Importance}
Across all architectures and decision stages, the attribution results reveal a stable hierarchy of feature categories. Contribution Activity emerges as the most influential signal (mean normalized importance $\approx$ 0.24-0.31), followed by Issue \& Maintenance Responsiveness ($\approx$ 0.19-0.26). Community Dynamics and Pull Request Quality Assurance contribute moderately, while Release \& Evolution Metrics consistently exhibit the lowest importance ($\approx$ 0.08-0.12). Figure~\ref{fig:bar-plot} aggregates importance scores across all models, confirming strong cross-model consensus: sustainability stage predictions are driven primarily by sustained contribution patterns and responsive maintenance behavior rather than episodic milestones such as releases.

\paragraph{Temporal Attribution Patterns}
For models incorporating temporal sequences (Stage-2 Heavy), Integrated Gradients analysis reveals a pronounced recency effect. Features from the most recent 6-12 months exhibit approximately 2.8 times higher attribution than those from 18-24 months prior. Attention analysis further corroborates this pattern, with roughly 68\% of total attention mass concentrated on the most recent eight months. These findings indicate that current engagement dynamics are substantially more predictive than distant historical activity.

\paragraph{Tabular Feature Contributions}
Explainability analysis of MLP-based components (Stage-1, Stage-2 Light, and the Club-Federation Expert) shows that aggregated statistics are more informative than raw monthly values. Six-month means, trend slopes, and delta statistics consistently receive higher SHAP importance than individual observations, indicating that compact summaries of momentum and stability provide strong predictive signals even in the absence of full temporal modeling.

\paragraph{Ablation-Based Validation}
Targeted feature ablation confirms that attribution scores reflect causal importance. Removing Contribution Activity results in a 16 percentage-point drop in accuracy (94.08\% $\rightarrow$ 78.31\%), followed by Issue Responsiveness (12\% drop). In contrast, removing Release \& Evolution features yields only minor degradation (3\% drop). The ablation ranking strongly correlates with XAI-derived importance (Spearman $\rho = 0.94$, $p < 0.01$), validating the robustness of the attribution results.

\paragraph{Model-Specific Attribution Profiles}
Different components of the hierarchical pipeline emphasize complementary feature subsets. The Stage-1 classifier prioritizes contribution intensity and community engagement to separate moderate-activity projects. The Stage-2 Heavy model distributes importance more evenly, leveraging both temporal dynamics and aggregated statistics. The Stage-2 Light model emphasizes maintenance and review responsiveness, while the Club-Federation Expert relies heavily on community structure indicators such as contributor concentration and diversity.


    
%


\section{Discussion}

\subsection{Summarization and Implications}
This study advances OSS sustainability research by showing how hierarchical temporal modeling combined with explainable learning can support accurate and interpretable sustainability stage prediction. The findings have implications for theory, methodology, and practice in OSS analytics. Our results demonstrate the value of hierarchical classification for modeling imbalanced and multi-regime OSS lifecycle structures. Critically, by decomposing prediction into staged, semantically meaningful decisions, the pipeline enables specialized models to focus on distinct sustainability regimes. The strong performance on dominant stages, together with improved, though still limited, recognition of minority stages, suggests that hierarchical routing can partially mitigate class imbalance by better aligning model capacity with label structure. 

Open source software is not a uniform phenomena, making approaches like the one described in this research essential for scaling our understanding the antecedents of project success in an uneven world. The explainability analysis shows that model decisions are grounded in domain-relevant signals rather than spurious correlations. Across architectures and decision components, the models consistently prioritize sustained contribution activity, issue responsiveness, and community engagement, factors widely recognized as central to OSS health \cite{eghbal2016roads}. The convergence of evidence across XAI methods provides confidence in these findings. In addition, the observed temporal decay pattern, where recent activity dominates predictive influence, aligns with established understandings of OSS evolution: near-term engagement is more indicative of sustainability than distant historical activity.

Our findings provide essential, useful insight for groups and organizations working to understand how to ensure the security, sustainability, and utility of Open Source Software. Sustained contribution activity and prompt issue responsiveness emerge as the strongest signals for improving sustainability, even when issue resolution itself is delayed. Community structure also plays a critical role: projects with low bus factor, high contributor turnover, or limited repeat participation face elevated sustainability risks, particularly when distinguishing between \texttt{club} and \texttt{federation} regimes. In contrast, release frequency appears less predictive, challenging traditional heuristics that emphasize milestone-driven development and suggesting that continuous contribution, engagement, and maintenance behaviors provide more reliable early indicators of project health. 

At the ecosystem level, the pipeline supports scalable, evidence-based monitoring. Reliable identification of \texttt{toy} and \texttt{contribMid} projects enables large-scale assessment, while interpretable attributions allow stakeholders to understand why a repository occupies a given sustainability stage. However, the lower confidence associated with minority classes cautions against fully automated decision-making for high-stakes interventions, underscoring the continued importance of human oversight.

Methodologically, the results indicate that combining temporal sequences with engineered tabular summaries strikes an effective balance between expressiveness and interpretability. Temporal models capture dynamic patterns that static features miss, while aggregated statistics provide stable, compact signals that generalize well across architectures. Together, these insights position the proposed framework as a meaningful step toward scalable, transparent, and theory-aligned OSS sustainability assessment.

\subsection{Limitations and Future Directions}

Despite the strong overall performance of the proposed hierarchical pipeline, several limitations constrain its predictive accuracy and generalizability. The most significant challenge arises from severe class imbalance: the minority sustainability stages (\texttt{club} and \texttt{federation}) account for fewer than 15\% of the dataset, limiting the availability of representative training examples and biasing learning toward majority classes. As a result, these stages exhibit lower precision and recall, reflecting the difficulty of capturing subtle and overlapping behavioral patterns from sparse observational data. This finding underscores an inherent challenge in OSS sustainability research: rare but strategically important project types remain difficult to model using observational repository data alone. This finding points to the need for future hybrid approaches that combine temporal modeling with richer governance, organizational, or funding signals.

The hierarchical design, while effective for structured decision-making, also introduces error propagation. Misclassifications in earlier stages may cascade to later decisions, disproportionately affecting underrepresented classes that require finer-grained distinctions. In addition, the model relies on repository-level activity traces that may contain noise or incomplete signals, particularly for governance or organizational dynamics that are not fully observable through standard OSS telemetry.

Several directions for future research follow naturally from these limitations. Advanced imbalance-aware strategies, such as temporal data augmentation, adaptive resampling, or confidence-calibrated ensemble weighting, may improve minority-class recognition. Incorporating complementary feature modalities, including code quality metrics, dependency networks, or governance signals, could further enhance discrimination between closely related sustainability stages. A key planned extension is evaluating the pipeline on larger and more diverse OSS ecosystems to assess cross-domain generalizability. Additionally, integrating real-time prediction into development platforms could enable continuous monitoring and appropriate intervention for projects showing early signs of sustainability risk.

\bibliographystyle{unsrt}  
\bibliography{references}

\appendix
\onecolumn
\section{Appendix A}

\setlength{\LTpre}{0pt}
\setlength{\LTpost}{0pt}

\begin{footnotesize}
\renewcommand{\arraystretch}{1.5}
\begin{longtable}{p{3.2cm} p{2.4cm} p{10.8cm}}
\caption{Illustration of the 20 activity metrics used in this study.}
\label{tab:feature-illustration}\\
\toprule
\textbf{Feature Category} & \textbf{Activity Metric} & \textbf{Description} \\
\midrule
\endfirsthead

\toprule
\textbf{Feature Category} & \textbf{Activity Metric} & \textbf{Description} \\
\midrule
\endhead

\midrule
\multicolumn{3}{r}{\emph{Continued on next page}} \\
\endfoot

\bottomrule
\endlastfoot

\multirow{7}{*}{\shortstack{Contribution activity\\metrics}} &
  Commit count &
  The total number of commits recorded in the repository during a given month. This metric captures overall development activity and code contribution intensity. \\ \cline{2-3} 
 &
  Committer count &
  The number of unique contributors who authored at least one commit within the month. This reflects the size of the active development community. \\ \cline{2-3} 
 &
  Pull requests &
  The total number of pull requests opened in the repository during the month, representing structured contribution activity beyond direct commits. \\ \cline{2-3} 
 &
  PR average commits &
  The average number of commits per pull request, indicating the typical size and complexity of proposed code changes. \\ \cline{2-3} 
 &
  PR average files &
  The average number of files modified per pull request, capturing the scope of individual contributions. \\ \cline{2-3} 
 &
  PR total files &
  The total number of files modified across all pull requests in the month, reflecting cumulative code churn associated with PR activity. \\ \cline{2-3} 
 &
  Bot contributors &
  The number of automated accounts (e.g., bots) contributing commits or pull requests during the month. This metric captures the extent of automation in development workflows. \\ \hline
\multirow{3}{*}{\shortstack{Community \&\\contributor dynamics}} &
  New contributors &
  The number of contributors making their first recorded contribution to the repository during the month, reflecting newcomer onboarding and community growth. \\ \cline{2-3} 
 &
  Repeat contributors &
  The number of contributors who have contributed in both the current and previous observation windows, indicating contributor retention and sustained engagement. \\ \cline{2-3} 
 &
  Bus factor &
  An estimate of contributor concentration, defined as the minimum number of contributors responsible for a large proportion of commits (e.g., 50-80\%). A lower bus factor indicates higher dependency on a small core team and greater sustainability risk. \\ \hline
\multirow{4}{*}{\shortstack{Issue \& maintenance\\responsiveness}} &
  Issues count &
  The number of issues opened during the month, reflecting reported bugs, feature requests, and maintenance demands. \\ \cline{2-3} 
 &
  Issue comments &
  The total number of comments posted on issues during the month, capturing discussion intensity and maintenance engagement. \\ \cline{2-3} 
 &
  Issue duration &
  The average time (in days) between issue creation and closure for issues resolved during the month, indicating maintenance efficiency. \\ \cline{2-3} 
 &
  Issue ttfr &
  The average time elapsed between issue creation and the first maintainer or contributor response, measuring responsiveness to reported problems. \\ \hline
\multirow{5}{*}{\shortstack{Pull request review \&\\quality assurance}} &
  Pr review comments &
  The total number of review comments posted on pull requests during the month, reflecting code review activity and collaborative quality control. \\ \cline{2-3} 
 &
  Pr review duration &
  The average time (in days) between pull request submission and final decision (merge or close), capturing review efficiency. \\ \cline{2-3} 
 &
  Pr total comments &
  The total number of comments (including review and discussion comments) across all pull requests in the month, indicating collaboration intensity. \\ \cline{2-3} 
 &
  Pr ttfr &
  The average time elapsed between pull request submission and the first review or comment, measuring responsiveness to code contributions. \\ \cline{2-3} 
 &
  Pr acceptance rate &
  The proportion of pull requests that are merged among all closed pull requests during the month, serving as an indicator of contribution quality and review selectivity. \\ \hline
Release \& evolution metrics &
  Releases &
  The number of official releases published during the month. This metric represents milestone-driven development and versioning activity.
\label{tab:feature-illustration}
\end{longtable}
\end{footnotesize}

\twocolumn

\section{Appendix B}

\subsection{Z-score Normalization} 
\label{subsec:z-score-norm}

\begin{equation}
\mathbf{X}_{i,t,f} = \frac{x_{i,t,f} - \mu_f}{\sigma_f + \epsilon}
\end{equation}

where $\mu_f$ and $\sigma_f$ are the mean and standard deviation across all projects and time steps for feature $f$, and $\epsilon = 10^{-8}$ prevents division by zero.

\subsection{Delta Temporal Features} \label{subsec:temporal-features}

\begin{equation}
\Delta \mathbf{X}_{i,t,f} = 
\begin{cases} 
0 & t = 0 \\
\mathbf{X}_{i,t,f} - \mathbf{X}_{i,t-1,f} & t > 0 
\end{cases}
\end{equation}

The final sequence tensor concatenates levels and deltas: $\mathbf{X}_i = [\mathbf{X}_i^\text{levels}, \Delta \mathbf{X}_i] \in \mathbb{R}^{T \times 2F}$.

\section{Appendix C}

\subsection{Detailed Model Architectures}

\begin{figure*}[t]
  \centering
  \includegraphics[width=\textwidth]{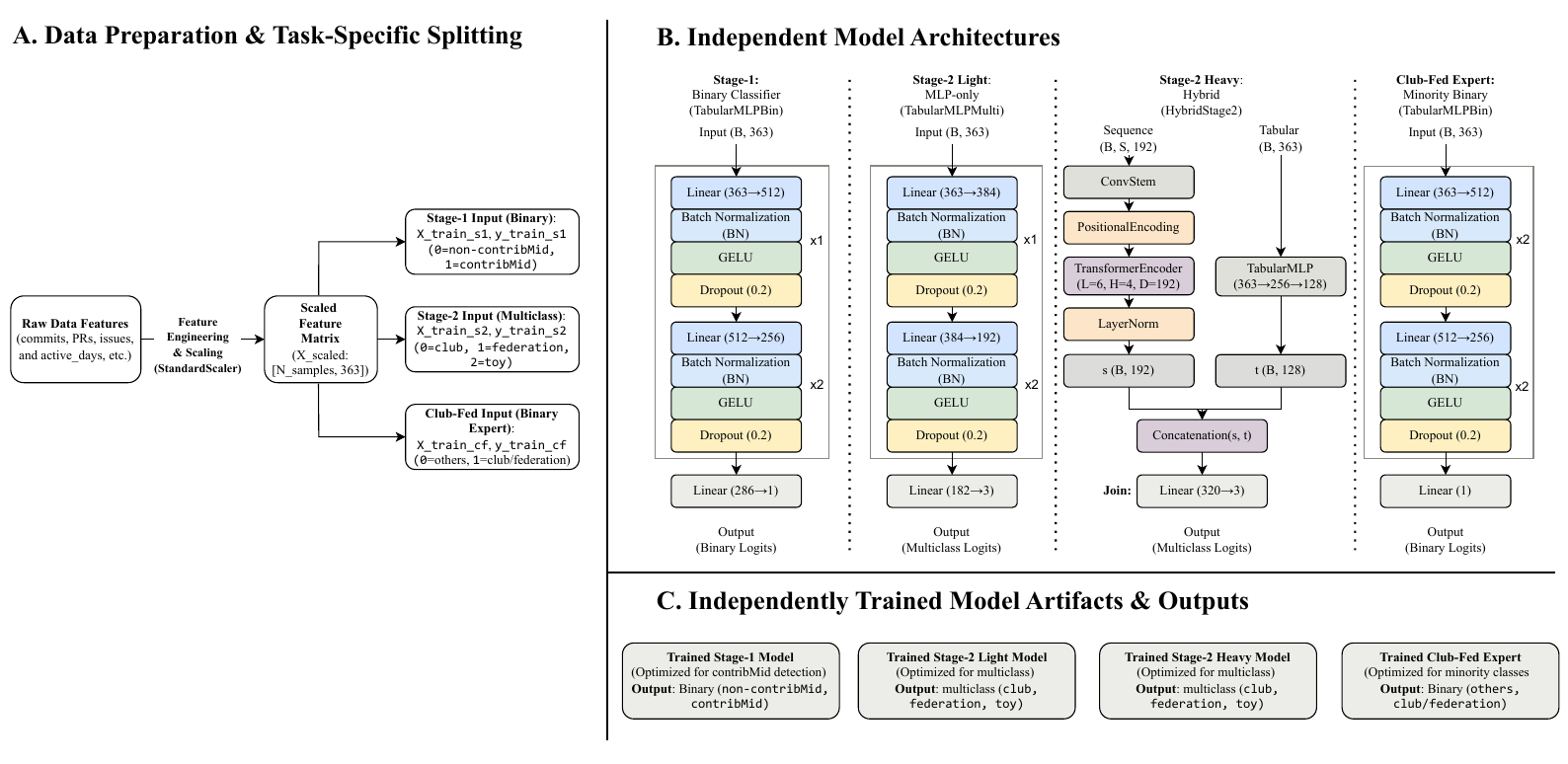}
  \caption{Detailed architectures of the four independently trained models. Panel A illustrates task-specific data preparation and splitting. Panel B shows the internal network architectures, including layer types and dimensions. Panel C summarizes the trained model artifacts and their respective outputs.}
  \label{fig:model_architectures}
\end{figure*}

Figure ~\ref{fig:model_architectures} shows detailed information regarding the each model or component within the hierarchical pipeline.

\subsection{Mathematical Formula to Integrate the Pipeline}

\subsubsection{Binary MLP Model}

\begin{equation}
\mathbf{h}^{(1)} = \text{MLP}^{(1)}(\mathbf{x}_i) \in \mathbb{R}^{H_1}
\end{equation}

\begin{equation}
\hat{y}^{(1)}_i = \sigma(\mathbf{w}^{(1)\top} \mathbf{h}^{(1)} + b^{(1)})
\end{equation}

where $\sigma$ is the sigmoid function, and the loss is focal loss to handle class imbalance:

\begin{equation}
\mathcal{L}_\text{focal} = -\alpha (1 - p_t)^\gamma \log(p_t)
\end{equation}

with $\alpha = 0.25$, $\gamma = 2.0$, and $p_t$ being the predicted probability for the true class \cite{lin2017focal}.

\subsubsection{Stage-2: Heavy model (Transformer + MLP}
\begin{equation}
\mathbf{z}_i^\text{seq} = \text{Transformer}(\mathbf{X}_i) \in \mathbb{R}^{D_\text{model}}
\end{equation}

\begin{equation}
\mathbf{h}_i^\text{tab} = \text{MLP}^\text{tab}(\mathbf{x}_i) \in \mathbb{R}^{D_\text{hidden}}
\end{equation}

\begin{equation}
\mathbf{h}_i^\text{heavy} = [\mathbf{z}_i^\text{seq}, \mathbf{h}_i^\text{tab}]
\end{equation}

\begin{equation}
\hat{\mathbf{y}}_i^\text{heavy} = \text{softmax}(\mathbf{W}^\text{heavy} \mathbf{h}_i^\text{heavy} + \mathbf{b}^\text{heavy})
\end{equation}

\begin{equation}
\text{Attention}(\mathbf{Q}, \mathbf{K}, \mathbf{V}) = \text{softmax}\left(\frac{\mathbf{Q}\mathbf{K}^\top}{\sqrt{d_k}}\right) \mathbf{V}
\end{equation}

\subsubsection{Stage-2: light model}
\begin{equation}
\mathbf{h}_i^\text{light} = \text{MLP}^\text{light}(\mathbf{x}_i)
\end{equation}

\begin{equation}
\hat{\mathbf{y}}_i^\text{light} = \text{softmax}(\mathbf{W}^\text{light} \mathbf{h}_i^\text{light} + \mathbf{b}^\text{light})
\end{equation}

\subsubsection{Club-Fed Expert}

\begin{equation}
\hat{y}_i^\text{cf} = \sigma(\text{MLP}^\text{cf}(\mathbf{x}_i))
\end{equation}

Here, $\hat{y}_i^\text{cf}$ represents the probability that the project belongs to the \texttt{club} class versus \texttt{federation}, computed independently from the ensemble predictions.

\subsubsection{Loss Function}

\begin{equation}
\mathcal{L}_\text{multi} = -\sum_{c=1}^C w_c \alpha_c (1 - p_c)^\gamma \log(p_c)
\end{equation}

where $w_c$ are class weights computed as inverse frequency, $\alpha_c = 0.25$, and $\gamma = 2.0$ \cite{lin2017focal}.

\section{Appendix D}
\subsection{Model Hyperparameters}

Table ~\ref{tab:all-hyper} shows detailed informaiton regarding each model's hyperparameter settings.


\begin{table}
\centering
\caption{Model Hyperparameters (Summary): (1) AdamW represents weighted Adam optimizer; (2) ES represents early stopping; (3) Sig. represents Sigmoid.}
\label{tab:all-hyper}
\footnotesize
\setlength{\tabcolsep}{3pt}
\begin{tabular}{@{}lcccc@{}}
\toprule
\textbf{Param.} & \textbf{S1} & \textbf{S2-H} & \textbf{S2-L} & \textbf{CF} \\
\midrule
Optimizer        & AdamW & AdamW & AdamW & AdamW \\
Weight Decay     & $10^{-4}$ & $10^{-4}$ & $10^{-4}$ & $10^{-4}$ \\
Learning Rate    & $8e{-4}$ & $6e{-4}$ & $8e{-4}$ & $8e{-4}$ \\
Batch Size       & 256 & 256 & 256 & 256 \\
Epochs           & ES & ES & ES & ES \\
Hidden Layers    & 3 & 2 & 3 & 2 \\
Layer Sizes      & 256–128–64 & 128–64 & 256–128–64 & 128–64 \\
Transformer L.   & -- & 6 & -- & -- \\
Attention Heads  & -- & 4 & -- & -- \\
Embed Dim        & -- & 192 & -- & -- \\
Pos. Encoding    & -- & Sin. & -- & -- \\
Activation       & ReLU & ReLU & ReLU & ReLU \\
Output Act.      & Sig. & Softmax & Softmax & Sig. \\
Dropout          & 0.1 & 0.1 & 0.1 & 0.1 \\
Loss             & \multicolumn{4}{c}{Focal ($\alpha=0.25,\gamma=2$)} \\
\bottomrule
\end{tabular}
\end{table}

\section{Appendix E}

\paragraph{Accuracy}
Accuracy measures the proportion of correctly classified instances:

\begin{equation}
\text{Accuracy} = \frac{TP + TN}{TP + TN + FP + FN}
\end{equation}

where $TP$, $TN$, $FP$, and $FN$ represent true positives, true negatives, false positives, and false negatives, respectively. For multi-class problems, this extends to all classes.

\paragraph{Balanced Accuracy}
Given the class imbalance in our dataset, we use balanced accuracy to account for performance across all classes equally:

\begin{equation}
\text{Balanced Accuracy} = \frac{1}{C} \sum_{c=1}^C \frac{TP_c + TN_c}{TP_c + TN_c + FP_c + FN_c}
\end{equation}

where $C = 4$ is the number of sustainability stages.

\paragraph{F1-Score Variants}
We compute both macro and weighted F1-scores to evaluate precision and recall trade-offs:

\begin{equation}
\text{Precision}_c = \frac{TP_c}{TP_c + FP_c}, \quad \text{Recall}_c = \frac{TP_c}{TP_c + FN_c}
\end{equation}

\begin{equation}
\text{F1}_c = 2 \cdot \frac{\text{Precision}_c \cdot \text{Recall}_c}{\text{Precision}_c + \text{Recall}_c}
\end{equation}

\begin{equation}
\text{Macro F1} = \frac{1}{C} \sum_{c=1}^C \text{F1}_c, \quad \text{Weighted F1} = \sum_{c=1}^C w_c \cdot \text{F1}_c
\end{equation}

where $w_c = n_c / N$ is the proportion of class $c$ in the test set.



\paragraph{Confidence Calibration}
For ensemble routing, we evaluate confidence calibration using expected calibration error (ECE):

\begin{equation}
\text{ECE} = \sum_{m=1}^M \frac{|B_m|}{N} \left| \text{acc}(B_m) - \text{conf}(B_m) \right|
\end{equation}

where $B_m$ are confidence bins, $\text{acc}(B_m)$ is accuracy in bin $m$, and $\text{conf}(B_m)$ is average confidence \cite{guo2017calibration}.

\section{Appendix F}

\subsection{Explainability Analysis}
\label{app:XAI}

\paragraph{Gradient-Based Attribution}
\begin{equation}
I_f = \left| \frac{\partial \mathcal{L}}{\partial x_f} \cdot x_f \right|
\end{equation}

\paragraph{SHAP}
\begin{equation}
\phi_i = \sum_{S \subseteq F \setminus \{i\}} \frac{|S|!(|F|-|S|-1)!}{|F|!} [f(S \cup \{i\}) - f(S)]
\end{equation}

where $F$ is the set of all features, $S$ is a subset of features, and $f$ is the model prediction function \cite{lundberg2017unified}.

\paragraph{Integrated Gradients}

\begin{equation}
IG_i(x) = (x_i - x'_i) \int_0^1 \frac{\partial F(x' + t(x - x'))}{\partial x_i} dt
\end{equation}

where $x'$ is the baseline input (typically zero), $x$ is the actual input, and $F$ is the model \cite{sundararajan2017axiomatic}.

\paragraph{Attention Validation}

\begin{equation}
\alpha_{ij}^h = \frac{\exp(e_{ij}^h)}{\sum_{k=1}^T \exp(e_{ik}^h)}
\end{equation}

where $\alpha_{ij}^h$ is the attention weight from position $i$ to $j$ in head $h$, and $e_{ij}^h$ is the attention logit \cite{vaswani2017attention}.

\end{document}